\documentclass[usegraphicx,usenatbib]{mn2e}

\newcommand{\xmm}{{\it XMM-Newton}}

\newcommand{\einstein}{{\it Einstein}}
\newcommand{\exosat}{{\it EXOSAT}}
\newcommand{\rosat}{{\it ROSAT}}

\newcommand{\beppo}{{\it BeppoSAX}}

\newcommand{\mekal}{{\sc mekal}}
\newcommand{\cemekl}{{\sc cemekl}}
\newcommand{\cevmkl}{{\sc cevmkl}}
\newcommand{\mkcflow}{{\sc mkcflow}}

\title{X-ray and UV observations of the dwarf nova VW Hyi in quiescence}

\author[Dirk Pandel, France A. C\'ordova and Steve B. Howell]
{Dirk Pandel$^1$, France A. C\'ordova$^2$ and Steve B. Howell$^2$\\
$^1$Department of Physics, University of California, Santa Barbara,
CA 93106, USA\\
$^2$Institute of Geophysics and Planetary Physics, Department of Physics,
University of California, Riverside, CA 92521, USA}

\pagerange{\pageref{firstpage}--\pageref{lastpage}}
\pubyear{2003}

\begin{document}

\maketitle

\label{firstpage}


\begin{abstract}

We present an analysis of X-ray and ultra-violet data of the dwarf nova VW Hyi 
that were obtained with \xmm\ during the quiescent state.
The X-ray spectrum indicates the presence of an optically thin plasma in the 
boundary layer that cools as it settles onto the white dwarf.
The plasma has a continuous temperature distribution that is well described by a 
power-law or a cooling flow model with a maximum temperature of 6--8$\:$keV.
We estimate from the X-ray spectrum a boundary layer luminosity of
$8\times10^{30}\:\mathrm{erg\:s^{-1}}$, which is only 20 per cent of the disk 
luminosity.
The rate of accretion onto the white dwarf is 
$5\times10^{-12}$M$_\odot\:$yr$^{-1}$, about half of the rate in the disk.
From the high-resolution X-ray spectra, we estimate that the X-ray emitting part 
of the boundary layer is rotating with a velocity of 
540$\:$km$\:$s$^{-1}$, which is close the rotation velocity of the white 
dwarf but significantly smaller than the Keplerian velocity.
We detect a 60-s quasi-periodic oscillation of the X-ray flux that is likely due 
to the rotation of the boundary layer.
The X-ray and the ultra-violet flux show strong variability on a time 
scale of $\sim$1500 s.
We find that the variability in the two bands is correlated and that the X-ray 
fluctuations are delayed by $\sim$100 s.
The correlation indicates that the variable ultra-violet flux is emitted near 
the transition region between the disk and the boundary layer and
that accretion rate fluctuations in this region are propagated to the X-ray
emitting part of the boundary layer within $\sim$100 s.
An orbital modulation of the X-ray flux suggests that the inner accretion disk 
is tilted with respect to the orbital plane.
The elemental abundances in the boundary layer are close to their solar values.

\end{abstract}



\section{Introduction}
\label{intro}

Dwarf novae are powerful X-ray sources with luminosities of 
$10^{30}$--$10^{33}\:\mathrm{erg\:s^{-1}}$.
The origin of these X-rays is thought to be the boundary layer between the 
white dwarf and the inner edge of the accretion disk.
There the disk material, initially moving at a Keplerian velocity, dissipates 
its kinetic energy while being decelerated to the rotation velocity of the white 
dwarf.
The structure of the boundary layer is still an unsolved problem in our 
understanding of equatorial accretion onto white dwarfs.
In the standard picture of dwarf novae, half of the accretion energy is radiated 
away by the disk, primarily at optical and ultra-violet (UV) wavelengths.
Unless the white dwarf is rotating rapidly, the other half of the energy should 
be liberated in the boundary layer and emitted as X-ray and extreme 
ultra-violet (EUV) photons.
However, observations have shown that for many dwarf novae the boundary layer 
luminosity is much lower than the disk luminosity.
In VW Hyi the X-ray flux during quiescence was found to be a factor of 4 lower 
than expected in the standard picture \citep{1991A&A...246L..44B}.

At a distance of 65 pc \citep{1987MNRAS.227...23W}, VW Hyi is one of 
the nearest and brightest cataclysmic variables.
It has been studied extensively at optical, UV, and X-ray wavelengths.
VW Hyi is a dwarf nova of type SU UMa and undergoes normal outbursts every 
$\sim$30 d and superoutbursts every $\sim$180 d.
It has an orbital period of 107 min and lies in a direction with an extremely 
low hydrogen column density $N_H=6\times10^{17}{\rm cm}^{-2}$ 
\citep{1990ApJ...356..211P}.
The X-ray spectrum in quiescence is consistent with emission from an optically 
thin, multi-temperature plasma with temperatures up to $\sim$12$\:$keV
\citep{1996A&A...307..137W}.
\citet{1987A&A...182..219V} detected coherent soft X-ray oscillations during 
superoutburst with a period of 14 s and suggested that they originate from the 
differentially rotating outer layers of the white dwarf or the boundary layer.
\citet{2001ApJ...561L.127S} inferred that the white dwarf's rotation 
velocity projected along the line of sight is 400--500 km$\:$s$^{-1}$,
corresponding to a rotation period of $\sim$80 s.
For our calculations, we assume an orbital inclination $i=60^\circ$ 
\citep{1998A&AS..129...83R}, a white dwarf mass $M_{wd}=0.63\:$M$_\odot$
\citep{1981A&A....97..185S}, and a white dwarf radius
$R_{wd}=8.3\times10^8\:$cm
\citep[derived from the mass-radius relationship in][]{1961ApJ...134..683H}.

In this paper we present our analysis of \xmm\ data obtained while VW Hyi 
was in a quiescent state.
With the high sensitivity of the X-ray detectors and their broad coverage of 
the X-ray band, we are able to determine the temperature distribution of the 
plasma in the boundary layer.
We search the X-ray and UV light curves for periodic and quasi-periodic 
oscillations and characterize the properties of the low-frequency noise 
(flickering).
The capability of \xmm\ for simultaneous X-ray and UV observations enables us to 
study the correlation between the variability of the boundary layer and the 
disk luminosity.
From the high-resolution RGS spectrum, we infer the rotational velocity of 
the X-ray emitting plasma in the boundary layer.
We derive abundances for elements with strong X-ray emission lines.


\section{Observations and data reduction}
\label{observations}

VW Hyi was observed with \xmm\ \citep{2001A&A...365L...1J} on 2001 October 19.
At the time of the observation, VW Hyi was in a quiescent state
22 d after a normal outburst and 23 d before a superoutburst
(derived from light curves provided by the AAVSO).
From the EPIC instruments \citep{2001A&A...365L..27T,2001A&A...365L..18S}
we obtained 18.7 ks (MOS) and 16.1 ks (PN) of continuous X-ray data.
The EPIC cameras were operating in small window mode (MOS) and in full frame 
mode (PN).
For MOS-1 and PN the medium blocking filters and for MOS-2 the thin blocking 
filter were selected.
The Optical Monitor \citep{2001A&A...365L..36M} performed 5 exposures with a 
total duration of 16.1 ks using the UVW1 filter (240--340 nm) and one exposure 
of 1.2 ks with the B filter.
The Optical Monitor (OM) was operated in fast mode with a sample time of 0.5 s.
We obtained 19.3 ks of data with the RGS \citep{2001A&A...365L...7D}.
Background levels were low during the entire observation.

From the EPIC MOS and PN data, we extracted source photons using a circular 
aperture with a radius of 40 arcsec.
The count rates given in this paper have been corrected for the 88 per cent
enclosed energy fraction of this aperture relative to a 60 arcsec aperture
(see {\it XMM-Newton Users' Handbook}).
The background rate, which was estimated from larger off-center regions,
contributes less than 1 per cent to the count rate in the source aperture.
We included in our analysis good photon events (FLAG=0) with patterns 0--12 for 
MOS and 0--4 for PN.
To create X-ray light curves, we applied a barycentric correction to the photon
arrival times and combined the events from all three EPIC instruments.
From the OM fast mode data, we extracted source photons using a circular 
aperture of 5 arcsec radius (81 per cent enclosed energy fraction).
Sky background contributed $\sim$2 per cent to the count rate in the source 
aperture.
The high UV brightness of VW Hyi caused significant coincidence losses at a 
level of 10--20 per cent.
We corrected for these coincidence losses using the method described in the 
{\it XMM-Newton Users' Handbook}.


\section{Timing analysis}
\label{timing}

\subsection{X-ray and UV light curves}
\label{xuvlight}

\begin{figure*}
\includegraphics{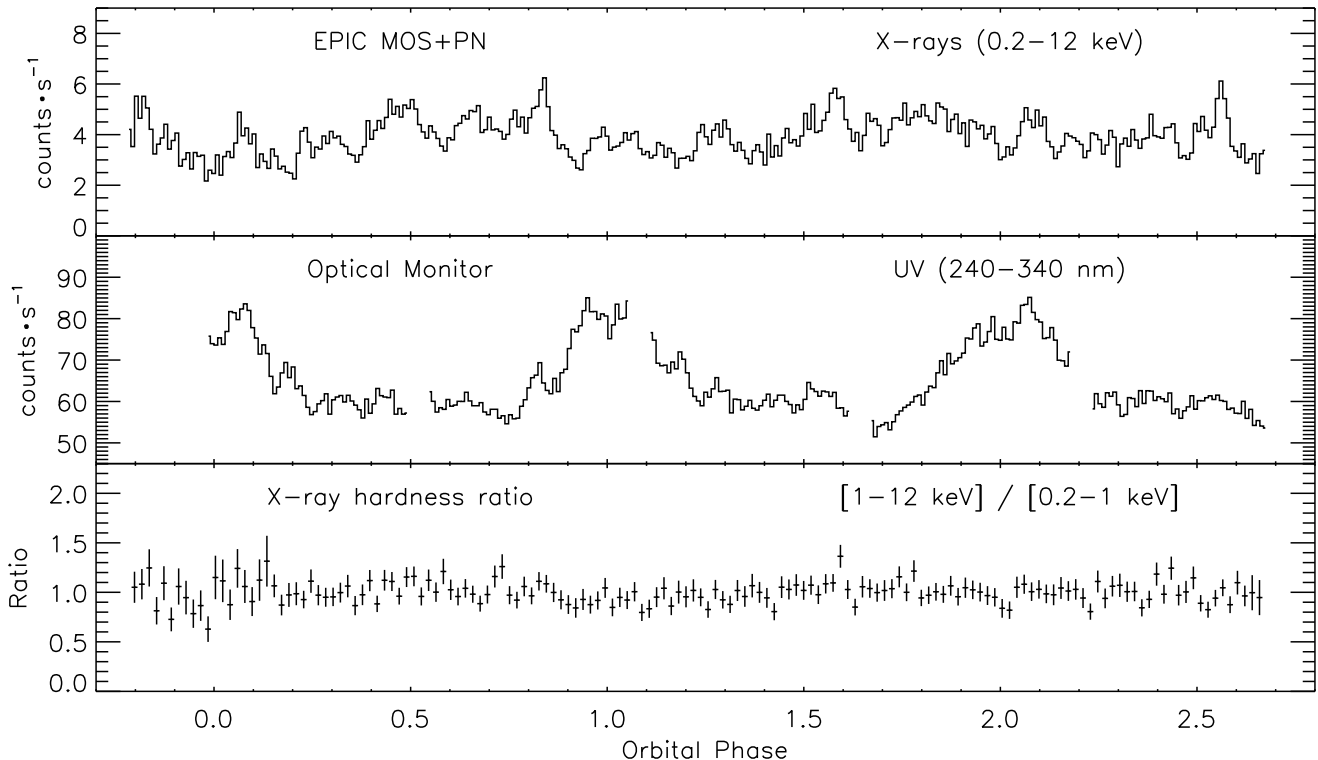}
\caption[X-ray and UV light curves]
{X-ray and UV light curves of VW Hyi obtained with \xmm\ on 2001 October 19
(see Section \ref{xuvlight}).
Also shown is the ratio of the count rates in the 1--12 keV and 0.2--1 keV
energy ranges.
VW Hyi has an orbital period of 107.0 min.
}
\label{lightcurve}
\end{figure*}

Fig. \ref{lightcurve} shows background subtracted X-ray and UV light curves 
covering three orbital cycles of the binary (orbital period 107.0 min).
As orbital phase 0 we selected BJD(TT) 2452201.748, which is the time of 
peak optical brightness as predicted by the ephemeris in \citet{1987MNRAS.225...93V}.
The ephemeris has an accumulated uncertainty of 250 s or 0.04 in phase.
With this choice of orbital phase 0, inferior conjunction of the secondary 
occurs at $\sim$0.27 \citep{1981A&A....97..185S}.

\begin{figure}
\includegraphics{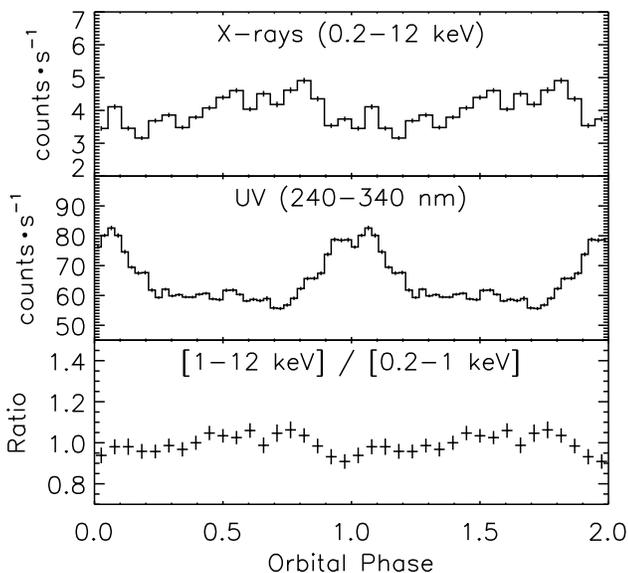}
\caption[Folded X-ray and UV light curves]
{X-ray/UV light curves and X-ray hardness ratio folded on the 
orbital period of 107.0 min.
}
\label{foldedlc}
\end{figure}

The X-ray light curves in Figs. \ref{lightcurve} and \ref{foldedlc} show
variability on two time scales.
Clearly visible is an orbital modulation of $\sim$30 per cent that is somewhat 
anti-correlated with the UV flux.
The hardness ratio in Fig. \ref{foldedlc} indicates that the X-ray spectrum
is softer during the fainter part of the orbital cycle.
The low-frequency flickering common in dwarf novae is seen as strong flaring 
with amplitudes up to $\sim$40 per cent and a recurrence time scale of 
$\sim$1500 s (see Section \ref{xuvcorr}).
Despite the strong variability of the X-ray flux on this time scale,
the hardness ratio in Fig. \ref{lightcurve} remains fairly constant.
The average count rate in the 0.2--12$\:$keV energy range is $4.0\:$s$^{-1}$.

The UV light curve in Figs. \ref{lightcurve} and \ref{foldedlc} show a strong
orbital modulation in the form of a broad hump.
This orbital hump is thought to be due to emission from the bright spot, the hot 
region where the accretion stream from the secondary impacts the outer 
edge of the disk \citep[e.g.][]{1995cvs..book.....W}.
On time scales shorter than the orbital period, the UV flux is less variable than
the X-ray flux (see Section \ref{xuvcorr} for a statistical analysis).
This is to be expected since the UV flux is dominated by emission from the inner
accretion disk and the white dwarf, while the X-rays mostly originate from the 
boundary layer.

In the UVW1 filter, a count rate of $1\:$s$^{-1}$ corresponds to a flux of
$4.5\times10^{-16}\:\mathrm{erg\:cm^{-2}\:s^{-1}}$\AA$^{-1}$ at 290 nm.
Accordingly, the UV flux from VW Hyi was varying between
$2.5$--$3.8\times10^{-14}\:\mathrm{erg\:cm^{-2}\:s^{-1}}$\AA$^{-1}$
with an average value of
$2.9\times10^{-14}\:\mathrm{erg\:cm^{-2}\:s^{-1}}$\AA$^{-1}$ or 8.1 mJy.
From the short B-band exposure (not shown), we estimate that the B magnitude was 
varying between 14.4 and 13.9.
Both the UV and the B-band brightness ranges agree with past measurements for VW 
Hyi during quiescence \citep{1987MNRAS.225...73P,1987MNRAS.225...93V}.

\subsection{Periodic and quasi-periodic oscillations}
\label{periodsearch}

\begin{figure*}
\includegraphics{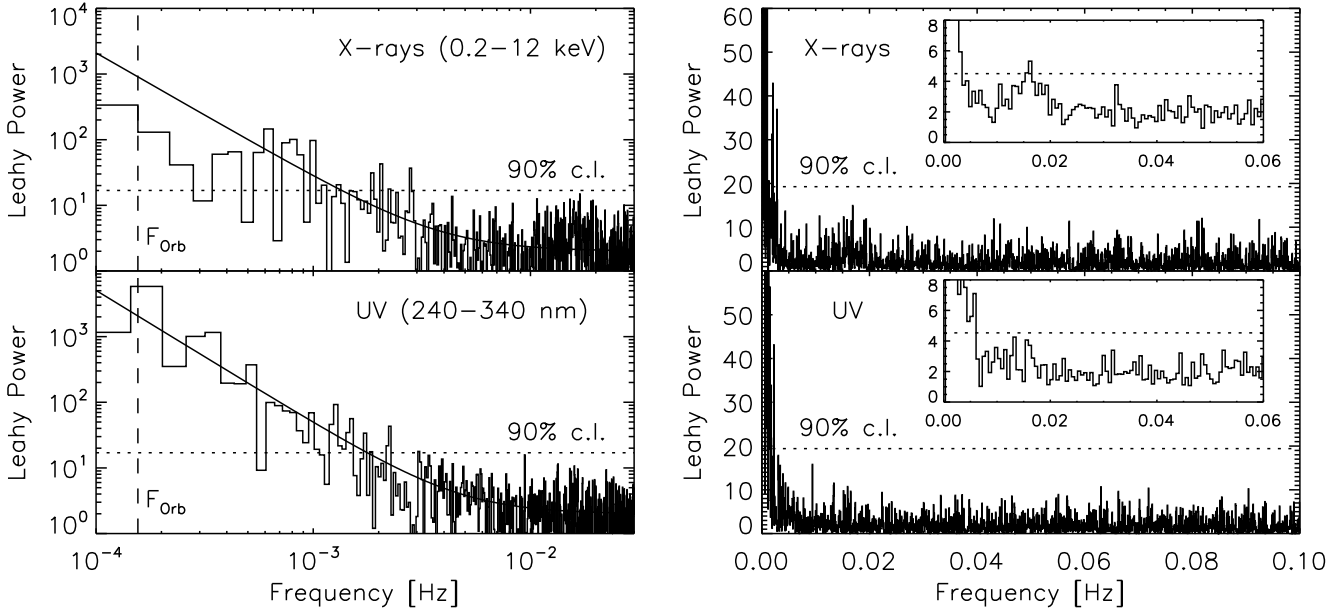}
\caption[X-ray and UV periodogram]
{Periodograms of the X-ray and UV light curves shown with logarithmic (left)
and linear (right) scales.
The spectral power is given in Leahy normalization for which each bin has an
average value of 2 provided that the source is non-variable and that the
observational noise is only due to photon counting statistics
(see Section \ref{periodsearch}).
The dotted lines indicate the 90 per cent confidence levels for the detection of 
strictly periodic signals and the dashed line marks the orbital frequency $F_{Orb}$.
The solid curves show the best fit of a power-law spectral density function. 
Averaged periodograms obtained by splitting the light curves into 10 equal time 
intervals are shown in the insets.
}
\label{fft}
\end{figure*}

Periodograms of the X-ray and UV light curves are shown in Fig. \ref{fft}.
The spectral power is given in Leahy normalization \citep{1983ApJ...266..160L},
which is suitable for data from photon counting detectors. 
In this normalization, the power in each spectral bin has a 
$\chi^2$-distribution with 2 degrees of freedom (mean 2 and variance 4),
provided that the source is non-variable.
The dotted lines in Fig. \ref{fft} are the 90 per cent confidence levels for the 
detection of strictly periodic signals.
For frequencies above 0.003 Hz, we do not find any evidence for periodic 
signals.
Although the detection threshold is exceeded below 0.003 Hz, the large spectral 
power is not due to periodic signals but rather the stochastic process of 
flickering common in dwarf novae (see Section \ref{xuvcorr}).
For the frequency range 0.005--1 Hz, we derived upper limits of 3.3 per cent and 
0.9 per cent, respectively, for the amplitude of any periodic signal in the 
X-ray and UV data (confidence level 90 per cent).

To search for quasi-periodic signals, we divided the light curves into shorter
time intervals and averaged the periodograms for all intervals.
The insets in Fig. \ref{fft} show the averages of 10 such periodograms.
For the X-ray data, there is a clear excess of spectral power between
0.01--0.02 Hz.  
Since the power is spread over many frequency bins, the excess is not due to a 
strictly periodic signal but rather indicates the presence of a quasi-periodic 
oscillation (QPO).
By summing the spectral power in the frequency range 0.011--0.021 Hz, we find that
the QPO is statistically significant at a $6\sigma$ level.
To estimate the parameters of the QPO, we fitted to the periodogram a spectral
density model consisting of a Gaussian plus a constant term for the photon counting
noise using a maximum likelihood method.
We obtained a center frequency of 0.016 Hz (62 s period), a FWHM frequency 
spread of 0.006 Hz (53--77 s period range), and an oscillation amplitude (RMS) 
of 6.2 per cent.
To determine the quality of the fit, we simulated spectra with the power density
function obtained from the fit by randomizing amplitudes and phases according
to the appropriate probability distributions.
We find that 96 per cent of the simulated power spectra yield a worse fit than
the \xmm\ data.

To further characterize the QPO, we assume that the peak in the power spectrum
is caused by a series of random pulses in the light curve that have a Gaussian
profile with a sinusoidal modulation.
From the above fit results, we can then derive some average properties of
these pulses.
We obtain a pulse duration of 110 s (FWHM of the Gaussian profile)
and $A^2_p/t_p=0.004\:$s$^{-1}$,
where $A_p$ is the pulse amplitude in counts$\cdot$s$^{-1}$
and $t_p$ the average pulse separation.
$A_p$ and $t_p$ cannot be determined independently, but the non-detection of 
individual pulses in the light curve requires $A_p\le1\:$s$^{-1}$
and therefore $t_p\le250\:$s.

The QPO is detected both for hard ($>$1$\:$keV) and soft ($<$1$\:$keV) X-rays
with no apparent difference in strength.
However, the QPO is not detected in the UV light curve, which places an upper 
limit of 0.7 per cent on the RMS amplitude in the 0.011--0.021 Hz range.
Our periodogram analysis did not reveal any other QPOs in the frequency range 
0.02--1 Hz.
This places an upper limit of 2.4 per cent for the X-ray data and 0.7 per cent 
for the UV data on the RMS amplitude of any QPO with a frequency spread of 
0.01 Hz.


\subsection{Flickering and X-ray/UV correlation}
\label{xuvcorr}

At low frequencies, the power spectrum is dominated by the flickering typical 
for dwarf novae.
This low-frequency variability is commonly modelled with a power-law spectral 
density function $S(f)\propto f^{-\beta}$
\citep[see e.g.][]{1995A&A...300..707T}.
The power-law index $\beta$ is indicative of the underlying process.
For example, random walk noise has an index $\beta=2$.
We fitted the periodograms with this power-law model plus a constant component
for the photon counting noise.
To avoid a bias from the orbital variations at a frequency of 0.000156 Hz,
we restricted the fit to frequencies above 0.0002 Hz.
The results of the fits are shown in Fig. \ref{fft} (solid curves).
We obtained for the power-law index $\beta$ a value of $1.90\pm0.19$ for the X-ray 
data and $2.02\pm0.14$ for the UV data.
Both indices are consistent with a random walk noise process.
To quantify the strength of the variability, we quote the RMS amplitude of the 
power-law noise above 0.0002 Hz.
We obtained an RMS of 18 per cent for the X-ray data and 7 per cent for the UV data.
To determine the quality of the fits, we simulated power spectra as described in
Section \ref{periodsearch}.
We found that in X-rays 45 per cent and in the UV 93 per cent of the simulated
spectra yield a worse fit than the \xmm\ data.
We investigated the dependence of the power-law parameters on the X-ray energy but could
not find statistically significant differences between hard and soft X-rays.

At frequencies below $\sim$0.0005 Hz, the X-ray spectral power appears to be lower than
predicted by the model fit, indicating that the power-law does not continue toward low
frequencies.
We determined a 99 per cent probability that the low spectral power in the range
0.0002--0.0005 Hz is not a coincidence and inconsistent with the power-law fit.
If frequencies below 0.0005 Hz are excluded from our fit, the power-law index
increases to $2.2\pm0.2$, which is still consistent with $\beta=2$.
The concentration of spectral power around 0.0007 Hz, which gives rise to the strong
variability on a $\sim$1500-s time scale (Fig. \ref{lightcurve}), may indicate the
presence of a QPO.
However, the duration of the \xmm\ observation is insufficient to distinguish
a QPO from a power-law spectral density function with a changing index $\beta$.

\begin{figure}
\includegraphics{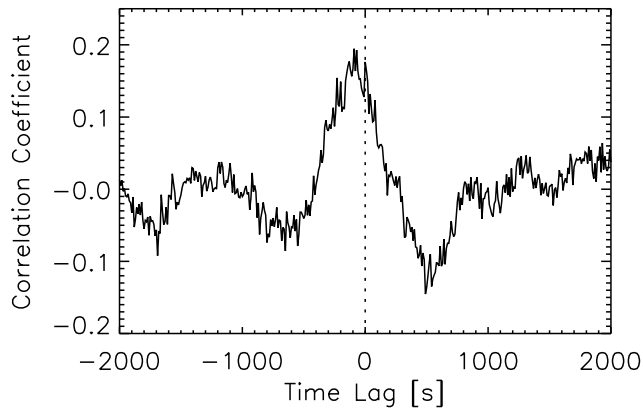}
\caption[X-ray/UV cross correlation]
{Cross-correlation between X-ray and UV light curves calculated after the 
variability on time scales longer than 2500$\:$s had been removed (see Section 
\ref{xuvcorr}).
The offset of the central peak with respect to 0 indicates that the X-ray 
variations lag behind those in the UV by $\sim$100 s.
}
\label{correlation}
\end{figure}

To investigate the connection between the X-ray and the UV variability,
we calculated the cross-correlation between the two light curves.
We removed the obvious correlation on the orbital time scale by first applying 
a high-pass filter with a cut-off frequency at 0.0004 Hz.
The resulting cross-correlation coefficient as a function of time lag is shown 
in Fig. \ref{correlation}.
A strong peak near zero time lag suggests a significant correlation between 
the X-ray and the UV variability.
The correlation coefficient is normalized such that it has a value of 1 for 
maximally correlated signals.
The high-frequency photon-counting noise introduces a bias that can reduce the 
correlation coefficient.
If we correct for this bias, the peak correlation coefficient increases to 0.5.
The central peak is offset with respect to zero time lag, indicating that the 
X-ray variations are delayed relative to those in the UV by $\sim$100 s.
The X-ray/UV correlation disappears when the cut-off frequency of the high-pass 
filter is increased to 0.001 Hz, demonstrating that the correlation is indeed 
due to the low-frequency flickering on the 1500-s time scale.
We also investigated the correlation between the hard and soft X-ray 
variability.
We found a strong correlation between the X-ray fluxes above and below 1$\:$keV,
but the time lag was consistent with zero to within a few seconds.


\section{Spectral analysis}
\label{spectral}

The medium resolution EPIC spectrum and the high resolution RGS spectrum are 
shown in Figs. \ref{epic} and \ref{rgs}.
Both spectra exhibit strong emission lines, indicating the presence of a hot, 
optically thin plasma.
The prominent lines identified by our model fits have been labelled.
Spectral fitting was performed with the XSPEC package version 11.2
\citep{1996adass...5...17A}.
As spectral models we used multi-temperature versions of the \mekal\ model,
which describes an optically thin, isothermal plasma based on calculations by 
\citet{1985A&AS...62..197M} and \citet{1995ApJ...438L.115L}.
For optimal fit results, we binned the EPIC and RGS spectra at one 
third of the respective FWHM detector resolution.
In addition, a minimum of 20 counts per bin were required for the EPIC spectra 
so that $\chi^2$-statistics could be used.
To retain the high resolution of the RGS spectra, we did not further rebin the 
RGS data but instead accounted for the low number of counts per bin by using 
C-statistics \citep{1979ApJ...228..939C}.


\subsection{EPIC spectrum}
\label{epicspec}

\begin{figure*}
\includegraphics{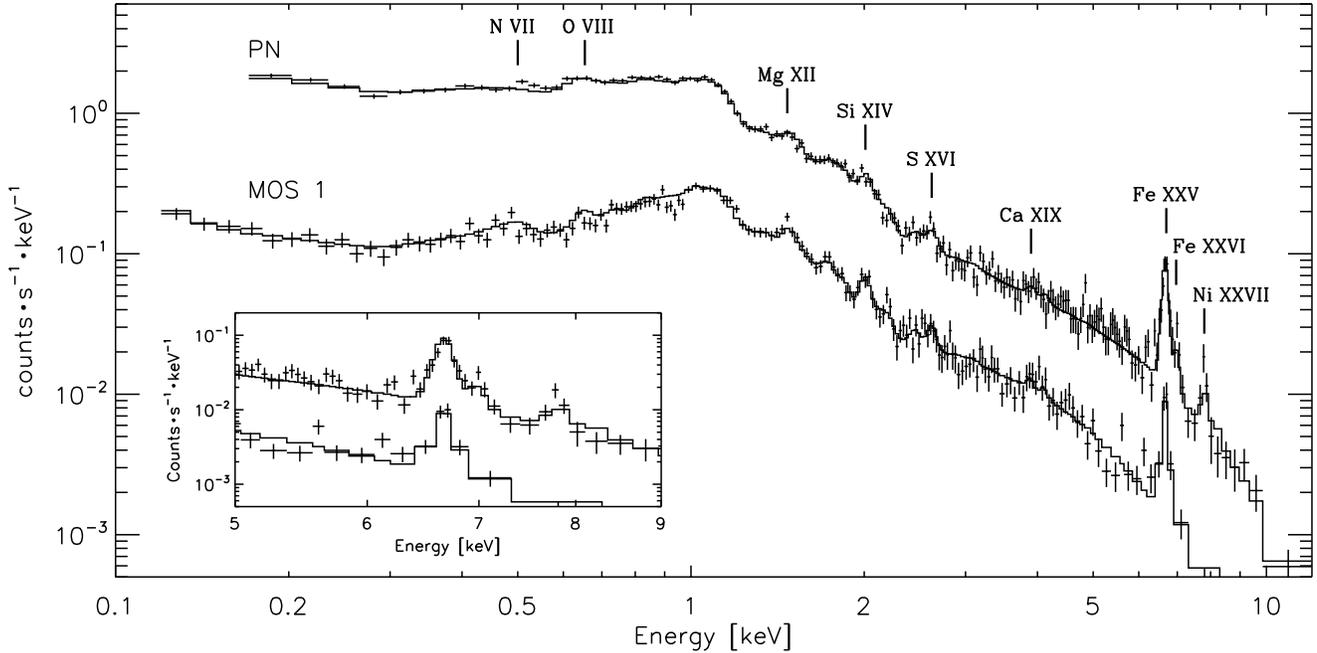}
\caption[EPIC spectrum]
{X-ray spectra obtained with the EPIC PN and MOS-1 cameras.
Prominent K-$\alpha$ emission lines have been labeled.
A magnified view of the Fe K-$\alpha$ line is shown in the inset.
For clarity, the MOS-1 data points have been shifted down by a factor of 2.
The solid curves show the best fit with the \cevmkl\ model
(see Section \ref{epicspec} and Table \ref{epicfits}).
}
\label{epic}
\end{figure*}

\begin{table*}
\caption[EPIC fit results]
{Results of our spectral fits of various optically thin, multi-temperature 
plasma models to the EPIC data (see Section \ref{epicspec}).
The \cemekl\ and \cevmkl\ models assume a power-law temperature distribution of 
the Emission Measure.
\mkcflow\ is a model for a cooling flow.
Except for the \cevmkl\ model, the abundance ratios were fixed at the solar 
values given in \citet{1989GeCoA..53..197A}.
For the latter three fits, the spectra from all three EPIC cameras were fit 
simultaneously.
In the \mkcflow\ model, the fitted low-temperature cut-off $kT_{min}$ was 0.08$\:$keV
(minimum allowed by model).
Uncertainties are given at a $1\sigma$ level.
}
\begin{center}
\begin{tabular}{lcccccccc}
\hline
Data sets & XSPEC & Power-law & $kT_{max}$ & Abundance &
\multicolumn{2}{c}{Flux [erg cm$^{-2}$ s$^{-1}$]}
& $\chi^2$ (d.o.f.) & Null hypothesis \\
& model & index $\alpha$ & [keV] & [solar] & 0.2--12 keV & bolometric & & 
probability \\
\hline
PN    & \cemekl & $1.24\pm0.04$ & $6.4\pm0.3$ & $1.03\pm0.05$ &
$6.4\times10^{-12}$ & $8.2\times10^{-12}$ & 212.1 (185) & 8.4 \% \\
MOS 1 & \cemekl & $1.63\pm0.09$ & $5.3\pm0.3$ & $0.99\pm0.08$ &
$6.1\times10^{-12}$ & $7.4\times10^{-12}$ & 280.6 (169) & 25.7 \% \\
MOS 2 & \cemekl & $1.38\pm0.05$ & $5.8\pm0.3$ & $1.05\pm0.08$ & 
$6.4\times10^{-12}$ & $8.0\times10^{-12}$ & 174.0 (171) & 42.2 \% \\
PN+MOS & \cemekl & $1.35\pm0.03$ & $6.0\pm0.2$ & $1.02\pm0.04$ & 
$6.3\times10^{-12}$ & $8.0\times10^{-12}$ & 605.3 (533) & 1.6 \% \\
PN+MOS & \mkcflow & -- & $7.8\pm0.2$ & $0.97\pm0.04$ &
$6.4\times10^{-12}$ & $7.8\times10^{-12}$ & 616.7 (533) & 0.7 \% \\
PN+MOS & \cevmkl & $1.28\pm0.04$ & $6.2\pm0.2$ & see Table \ref{abund} &
$6.3\times10^{-12}$ & $8.1\times10^{-12}$ & 557.3 (521) & 13.1 \% \\
\hline
\end{tabular}
\end{center}
\label{epicfits}
\end{table*}

The X-ray spectra of dwarf novae in quiescence are commonly fit with one or 
two single-temperature plasma models.
We attempted to fit these models to the EPIC spectra but found that they are 
inconsistent with the data.
However, a good fit could be achieved with a combination of three 
single-temperature \mekal\ models.
For the temperatures of the three components we found 0.5, 1.5, and 5$\:$keV.
Adding a fourth component to the model did not improve the fit.
It is evident from this result that the X-ray emitting plasma has a continuous 
temperature distribution, at least over the range 0.5--5$\:$keV.
The multi-temperature nature can be expected for accreting gas that cools as it 
settles onto the white dwarf.
To account for the spread in temperature, we attempted to fit several 
\mekal-based continuous-temperature models provided by XSPEC.

The best results were achieved with the \cemekl\ model, which fit the data
slightly better than the three-temperature model but with a smaller 
number of parameters.
As the simplest of the \mekal-based multi-temperature models, \cemekl\ assumes 
that the {\it Emission Measure} ($EM$) has a power-law temperature-dependence
\begin{equation}
\frac{d\,EM}{d\,\log T}=DEM(T_{max})\times\left(\frac{T}{T_{max}}\right)^\alpha
\end{equation}
where $T_{max}$ is the maximum temperature of the plasma and $DEM(T_{max})$ the 
{\it Differential Emission Measure} at $T_{max}$
\begin{equation}
DEM(T_{max})=\left(\frac{d\,EM}{d\,\log T}\right)_{T=T_{max}}
\end{equation}
The integration in the \cemekl\ code is performed logarithmically (note the 
$\log T$ above), so that a constant $EM(T)$ corresponds to $\alpha=1$ 
\citep{1997MNRAS.288..649D}.
The results of our fits are summarized in Table \ref{epicfits}.
Rows 1--3 show the parameters obtained when the spectrum from each of the three 
EPIC cameras was fit individually.
As can be seen in the null hypothesis probabilities, the \cemekl\ model is in 
very good agreement with the data.
The three EPIC spectra yield somewhat different values for the parameters 
$\alpha$ and $T_{max}$.
We attribute this to small discrepancies in the cross-calibration of the EPIC 
detectors.
Because the parameters $\alpha$ and $T_{max}$ are strongly anti-correlated, the 
differences between the three fits are not as significant as they might appear.
Row 4 in Table \ref{epicfits} shows the result for a simultaneous fit of the 
three EPIC spectra.
The null hypothesis probability for the combined fit is significantly lower than 
for the individual fits, which is a further indication for the cross-calibration 
discrepancies.
Assuming isotropic emission and a distance of 65 pc, we obtain from the fit a 
{\it Differential Emission Measure} $DEM(T_{max})=4.3\times10^{53}$cm$^{-3}$ at 
$kT_{max}=6.0\:$keV.
Integrating the power-law model from $T=0$ to $T_{max}$ yields a total
{\it Emission Measure} $EM=1.4\times10^{53}$cm$^{-3}$.
The neutral hydrogen column density $N_H$ toward VW Hyi is fairly low at
$6\times10^{17}{\rm cm}^{-2}$ \citep{1990ApJ...356..211P}.
Accordingly, adding photoelectric absorption to our model did not improve the 
fit.
We derived an upper limit for $N_H$ of $2.2\times10^{18}{\rm cm}^{-2}$
(at 90 \% c.l.).

In the \cemekl\ model, the elemental abundance ratios are fixed at solar values 
and only the total abundance is varied.
A significant improvement of the fit could be achieved with the \cevmkl\ model
(solid curve in Fig. \ref{epic} and row 6 in Table \ref{epicfits}).
In this model, the abundances for the 13 most common elements from C to Ni 
can be varied independently.
Table \ref{abund} shows the best-fit values and the 90 per cent confidence 
ranges of the abundances obtained with the \cevmkl\ model.
Good constraints are found for the elements O, Mg, Si, S, and Fe,
all of which have abundances near the solar values.
The abundances are mainly derived from the K-$\alpha$ emission line of each 
element.
The most prominent of these lines are labeled in Fig. \ref{epic}.
Because of the lower spectral resolution of the EPIC detectors, confusion of 
emission lines may bias the measurements for the less abundant elements.
In particular, the estimates for Ne and Na are unreliable, since their 
K-$\alpha$ lines lie in the range 0.7--1.3$\:$keV, which is dominated by Fe L-shell 
lines.

A good fit, though slightly worse than with the \cemekl\ model, was achieved 
with the \mkcflow\ model (row 5 in Table \ref{epicfits}).
\mkcflow\ is based on the cooling flow model by \citet{1988cfcg.work...53M} and 
assumes that the Emission Measure for each temperature is proportional to the 
inverse of the bolometric luminosity (i.e. proportional to the cooling time).
The variable parameters in the model are the minimum and maximum temperatures 
of the flow ($T_{min}$ and $T_{max}$), the elemental abundance (abundance ratios 
fixed at solar values), and the flux normalization.
The good agreement of the X-ray spectrum with the \mkcflow\ model indicates the 
presence of hot plasma in the boundary layer that cools as it settles onto the 
white dwarf.
However, the cooling flow model assumes constant pressure and spherical 
symmetry, which may not be valid in the strong gravitational field of the 
white dwarf and for the shear and turbulence in a rotating, belt-like boundary 
layer.
Since the \cevmkl\ model provides the best fit, we will use it in our further 
analysis.

The maximum plasma temperature of 6--8$\:$keV in VW Hyi is low compared to other 
dwarf novae.
For instance, \citet{2003ApJ...586L..77M} found maximum temperatures of $\sim$20$\:$keV for 
U Gem and $\sim$80$\:$keV for SS Cyg, while \citet{1997MNRAS.288..649D} obtained 
$\sim$20$\:$keV for SS Cyg.
From earlier observations of VW Hyi during quiescence, \citet{1999A&A...349..588H} and
\citet{1996A&A...307..137W} derived similarly low temperatures of 10$\:$keV and 11$\:$keV,
respectively, using a cooling flow model.
During these observations, however, the X-ray flux was about twice that observed with
\xmm\ (see below), which may explain our somewhat lower result of $7.8\:$keV.

According to our model fits, the bolometric flux from the boundary layer
is $8\times10^{-12}\:\mathrm{erg\:cm^{-2}\:s^{-1}}$.
Since most of the flux is emitted inside the \xmm\ detector band, this estimate 
is fairly independent of the spectral model (see Table \ref{epicfits}).
For a distance of 65 pc, the corresponding boundary layer luminosity is 
$L_{bl}=8\times10^{30}\:\mathrm{erg\:s^{-1}}$.
Here we assumed that, because of the high orbital inclination, half of the 
belt-like boundary layer is obscured by the white dwarf.
We ignored reflection of X-rays by the white dwarf and a possible obscuration of 
the boundary layer by the disk.
In comparison with previous observations during quiescence, the X-ray flux 
observed with \xmm\ was about half of that found with \beppo\ 
\citep{1999A&A...349..588H}, \rosat\ \citep{1991A&A...246L..44B}, and \exosat\ 
\citep{1987MNRAS.225..141V}, and about twice that seen with \einstein\ 
\citep{1981ApJ...245..609C}.
Despite the differences in X-ray flux, VW Hyi had a similar optical brightness 
during these observations.

To further investigate the orbital modulation of the X-ray flux (Figs. 
\ref{lightcurve} and \ref{foldedlc}), we extracted individual spectra for
the bright and the faint parts of the light curve
(orbital phases 0.4--0.9 and 0.9--0.4, respectively).
We find that the spectrum is softer during the faint phase and that the ratio 
of faint-phase to bright-phase spectrum is decreasing by $\sim$10 per cent from 
the lowest to the highest X-ray energies.
The softening of the spectrum can also be seen in the hardness ratio in Fig.
\ref{foldedlc}.
This trend is opposite of what would be expected if the decline in brightness 
were due to absorption by some intervening gas.
Possible causes for the orbital X-ray modulation are discussed in Section \ref{xraymod}.


\subsection{RGS spectrum}
\label{rgsspec}

\begin{figure*}
\includegraphics{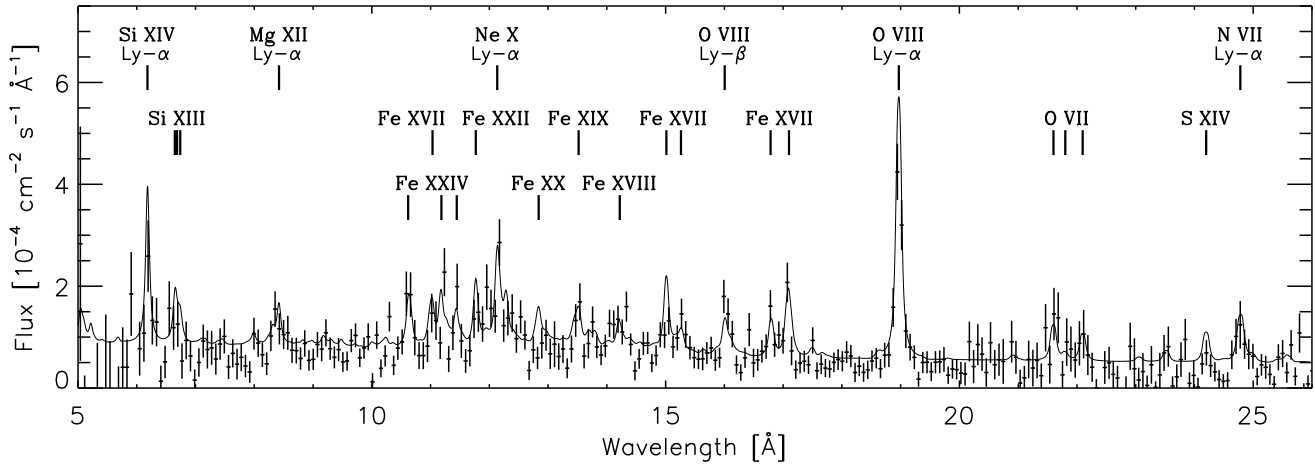}
\caption[RGS spectrum]
{Flux-calibrated RGS spectrum showing the combined $1^\mathrm{st}$ and 
$2^\mathrm{nd}$ order RGS-1 and RGS-2 data.
Prominent emission lines have been labeled.
The solid curve shows the best fit of the \cevmkl\ model
(see Section \ref{rgsspec} and Table \ref{epicfits}).
The data is binned at 0.072 \AA, the FWHM detector resolution.
}
\label{rgs}
\end{figure*}

\begin{table}
\caption[Abundances]
{Elemental abundances derived from our spectral fits with the \cevmkl\ 
model (see Sections \ref{epicspec} and \ref{rgsspec}).
Abundances are given relative to the solar values in 
\citet{1989GeCoA..53..197A}.
Shown are the best-fit values and the 90 per cent confidence ranges.
The more reliable abundance estimates have been highlighted.
}
\begin{center}
\begin{tabular}{lcccc}
\hline
Element & \multicolumn{2}{c}{EPIC} & \multicolumn{2}{c}{RGS} \\
        & best fit  & 90 \% c.l. & best fit & 90 \% c.l. \\
\hline
C   & 1.5   & 0.3--2.9    & {\bf 0.82}  & {\bf 0.52--1.65}  \\
N   & 1.1   & 0.0--2.4    & {\bf 1.6}   & {\bf 0.9--2.9}    \\
O   & {\bf 0.95}  & {\bf 0.79--1.12}  & {\bf 0.80}  & {\bf 0.67--0.91}  \\
Ne  & 0.24  & 0.0--0.55   & {\bf 0.81}  & {\bf 0.29--1.26}  \\
Na  & $>10$ & $>3.4$      & 0.0   & 0.0--10.4   \\
Mg  & {\bf 1.27}  & {\bf 0.91--1.65}  & 0.9   & 0.4--2.3    \\
Al  & 4.5   & 1.6--7.6    & 0.0   & 0.0--6.5    \\
Si  & {\bf 1.30}  & {\bf 1.05--1.57}  & 2.8   & 1.6--3.8    \\
S   & {\bf 1.10}  & {\bf 0.77--1.45}  & 1.9   & 0.9--2.9    \\
Ar  & 0.0   & 0.0--0.75   & 4.8   & 0.2--9.4    \\
Ca  & {\bf 1.8}   & {\bf 0.6--2.9}    & 0.0   & 0.0--3.0    \\
Fe  & {\bf 1.10}  & {\bf 0.99--1.22}  & {\bf 0.68}  & {\bf 0.59--0.77}  \\
Ni  & {\bf 2.0}   & {\bf 1.1--3.0}    & 0.0   & 0.0--1.1    \\
\hline
\end{tabular}
\end{center}
\label{abund}
\end{table}

The RGS spectrum (Fig. \ref{rgs}) shows a number of prominent emission lines, in 
particular the Lyman-$\alpha$ lines of some common elements and the Fe L-shell 
lines between 0.7 and 1.3$\:$keV.
As for the EPIC spectra, we fitted the RGS spectrum with the model for an 
optically thin plasma with a power-law temperature distribution (\cevmkl).
Because of the smaller energy range covered by the RGS, the model parameters 
$\alpha$ and $T_{max}$ were not well constrained.
We therefore fixed these parameters at the values obtained from the EPIC spectra 
($\alpha=1.28$ and $kT_{max}=6.2\:$keV) and only fit the elemental abundances.
The best-fit model is shown as a solid curve in Fig. \ref{rgs}.
The elemental abundances (Table \ref{abund}) are in general agreement with those 
obtained from the EPIC spectra.
The RGS spectrum tends to yield good estimates for elements with prominent 
Lyman-$\alpha$ lines in the RGS detector band (C, O, Ne).
For other elements (Mg, Si, S), the Lyman-$\alpha$ lines are fairly weak and 
the more sensitive EPIC detectors provide better constraints.
In Table \ref{abund} the more reliable abundance estimates have been 
highlighted.
For most elements, our results are consistent with near solar abundances.
We cannot confirm the large over-abundance of C and Si found by 
\citet{1996AJ....111.2386H} for a UV emitting ring rotating with the Keplerian 
velocity.
However, the UV spectra on which these results are based were obtained 10 days 
after a superoutburst and the ring may actually be a post-outburst equatorial 
belt on the white dwarf \citep{2002MNRAS.335...84W}.
The abundances in such a belt may be enhanced due to mixing with the outer 
layers of the white dwarf.
In contrast, our estimates derived from the X-ray spectrum provide the 
abundances in the accreting gas.

Good estimates from both the RGS and the EPIC spectra were obtained only for the 
abundances of oxygen and iron.
While for oxygen the two values are consistent, there is a noticeable discrepancy 
for iron.
In the EPIC spectrum, the iron abundance is mainly derived from the Fe {\sc xxv} 
K-$\alpha$ line at 6.7$\:$keV, while in the RGS spectrum the L-shell lines of Fe 
{\sc xvii--xxiv} determine the abundance.
We interpret the discrepancy for iron as an indication that the power-law model 
slightly overestimates the emission measure for temperatures at which iron is in 
the ionization states Fe {\sc xvii--xxiv}.
The model appears to overpredict the strength of the Fe {\sc xx} line at 12.8 
\AA\ and the Fe {\sc xvii} line at 15.0 \AA.
The cause for this discrepancy may be a high opacity for these lines.
According to \citet{1999ApJ...516..482B}, the Fe {\sc xvii} line at 15.0 \AA\ 
weakens significantly relative to the other visible Fe {\sc xvii} lines for 
column densities $>$$10^{17}\:$cm$^{-2}$.
For the accretion rate determined in Section \ref{lumin} and for a boundary 
layer covering 10 per cent of the white dwarf, such a column density can be 
present if the iron in the cooling gas remains in the Fe {\sc xvii} ionization 
state for $\sim$5 s.

The width of the O {\sc viii} line at 19.0 \AA\ is slightly larger than the 
spectral resolution of the RGS.
The likely cause is Doppler-broadening due to the rotation of the boundary layer 
around the white dwarf.
Broadening due to the orbital motion of the white dwarf itself is about one 
order of magnitude smaller and can be neglected.
We fitted the line profile assuming that the X-ray emitting part of the boundary
layer is a thin, equatorial belt near the white dwarf's surface and found a significantly
non-zero rotation velocity $V_{bl}\sin{i}=520\:$km$\:$s$^{-1}$ with a 90 per cent confidence
range of 80--810 km$\:$s$^{-1}$.
Here $i$ is the orbital inclination of the binary.
A slightly better constraint can be obtained by fitting the entire RGS spectrum 
as described earlier in this section but with the inclusion of the Doppler-broadening.
We find a best-fit velocity $V_{bl}\sin{i}=540\:$km$\:$s$^{-1}$ with a 90 per cent 
confidence range of 280--810 km$\:$s$^{-1}$.
This result is in agreement with the rotation velocity of 750$\:$km$\:$s$^{-1}$
derived from the 60-s oscillation (see Section \ref{oscillation}).


\section{Discussion}

\subsection{60-s X-ray oscillation and boundary layer rotation}
\label{oscillation}

Coherent dwarf nova oscillations (DNOs) and quasi-periodic oscillations (QPOs)
have been observed in many dwarf novae at X-ray, ultra-violet and optical 
wavelength \citep{1995cvs..book.....W}.
In VW Hyi the only X-ray oscillation previously found was a coherent 14-s
DNO during superoutburst \citep{1987A&A...182..219V}.
As its origin, the authors suggested the differentially rotating outer layers of 
the white dwarf or the boundary layer.
At optical wavelength, DNOs have been observed in the period range 20--36 s and
QPOs at the periods 23 s, 88 s, 253 s, and 413 s
\citep[see Table 8.2 in][]{1995cvs..book.....W}.
Recently, \citet{2002MNRAS.333..411W} found DNOs with a period evolving 
thoughout the outburst from 14 s to $>$40 s and QPOs with evolving periods of 
hundreds of seconds.
In all cases, the oscillations were seen during or shortly after an outburst.

The 60-s X-ray QPO we found with \xmm\ is the first oscillation observed in 
VW Hyi during quiescence.
The QPO is detected with equal strength in the soft and hard X-ray bands (i.e. 
the spectral slope is similar to that of the overall X-ray spectrum), which 
indicates that the oscillation is caused by small brightness variations in the 
boundary layer.
Emission from a magnetic pole can be ruled out since the oscillation is not 
strictly periodic.
The 60-s oscillation may be related to the rotation of the X-ray emitting gas in 
the boundary layer.
Since VW Hyi is a high-inclination system ($i=60^\circ$), intermittently 
brightened parts of the boundary layer may be periodically obscured as they 
rotate around the white dwarf, thus creating a weakly coherent oscillation.
The average live-time of the brightness fluctuations is $\sim$110$\:$s
(see Section \ref{periodsearch}), similar to the coherence time of the QPO.

Assuming that the 60-s oscillation is due to rotation around the white dwarf, we 
can estimate the rotation velocity of the X-ray emitting gas in the boundary layer.
For an orbital inclination $i=60^\circ$ and a radius equal to the white dwarf 
radius $R_{wd}=8.3\times10^8\:$cm, we find a rotation velocity 
$V_{bl}\sin{i}=750\:$km$\:$s$^{-1}$.
This is slightly larger than the velocity of 540$\:$km$\:$s$^{-1}$ we determined from
the line-broadening (Section \ref{rgsspec}) but still within the 90 per cent
confidence range.
The velocity derived from the QPO may be biased since it depends on the somewhat
uncertain values of $i$ and $R_{wd}$.
Our velocity measurements indicate that the boundary layer is rotating significantly
slower than the inner accretion disk, which has a Keplerian velocity of 
$\sim$$3200\:$km$\:$s$^{-1}$.
However, the boundary layer velocity is close to the rotation velocity of the
white dwarf $V_{wd}\sin{i}=400$--$500\:$km$\:$s$^{-1}$ determined by 
\citet{2001ApJ...561L.127S} from the width of UV-detected atmospheric lines.
It appears that the accreting gas dissipates most of its rotational kinetic 
energy and decelerates to almost the white dwarf's rotation velocity before 
it can radiate efficiently in X-rays.


\subsection{X-ray/UV correlation}
\label{correl}

As discussed in Section \ref{xuvcorr}, we found that the variability of the 
X-ray and UV flux seen on a time scale of $\sim$1500$\:$s is correlated and that 
the X-ray flux variations are delayed by $\sim$100$\:$s.
In the past, correlation studies have been rare because of the technical 
difficulties of simultaneous X-ray and optical/UV observations.
With the Optical Monitor on board \xmm, the search for correlations is now
greatly simplified.
A similar correlation as the one in VW Hyi was found by 
\citet{1983ApJ...270..211J} in the nova-like variable TT Ari.
There the X-ray variability was also delayed by $\sim$1 min.

The observed correlation is somewhat surprising since the X-ray and UV emission 
originate from distinct regions in the binary.
While the source of the X-ray emission is the boundary layer very close to the 
white dwarf's surface, the UV flux is emitted by the inner accretion disk 
\citep[e.g.][]{1989A&A...211..131L} and the white dwarf.
Because of the relatively low temperatures, the disk or white dwarf do not 
contribute to the X-ray flux.
Similarly, the contribution of the boundary layer to the UV flux, as 
extrapolated from the X-ray spectrum, is only $\sim$$3\times10^{-4}$ of the 
total flux.

Because the X-ray flux variations are trailing those in the UV, we can rule out 
reprocessing of X-rays in the accretion disk as the cause of the correlation.
Further investigations of the X-ray data did not revealed noticeable differences 
between the spectrum during flares and the spectrum between flares.
We therefore suspect that the X-ray variability is caused by fluctuations of the 
accretion rate onto the white dwarf.
These fluctuations may be due to accretion rate variations in the disk, directly 
seen via the variable UV flux, that propagate inward to the boundary layer.
However, the viscous time scale, i.e. the time scale for matter to diffuse 
inward through the disk, is much longer than the observed delay of 
$\sim$100$\:$s \citep[e.g.][]{1992apa..book.....F}.
The variable UV flux must therefore be emitted from near the transition region 
between the disk and the boundary layer.
There the accreting gas is decelerated well below the Keplerian velocity and is 
no longer supported by rotation against radial in-fall.
Accretion rate fluctuations near the inner disk edge may be quickly propagated
to the inner parts of the boundary layer.
\citet{1994A&A...288..175M} suggested that, during quiescence, accretion onto the 
white dwarf does not occur directly from the disk but via a corona that is 
formed by evaporation of the inner disk.
In this model, the UV variability may be connected to variations of the 
evaporation rate that modulate the amount of gas in the corona and, with some 
time delay, the accretion rate onto the white dwarf.

As a feasibility test, we estimate the minimum size of the region responsible 
for the variable UV flux.
We assume that the region is emitting as a blackbody and that it is causing the 
7 per cent variability of the flux in the UVW1 filter (see Section \ref{xuvcorr}).
A further constraint on the blackbody can be derived from the non-detection 
of UV emission at wavelength shorter than 98 nm \citep{1996ApJ...466..964L}.
We find a maximum temperature of 3 eV and a minimum emitting area of 
$6\times10^{16}\:$cm$^2$.
The corresponding width of an annular region with a radius equal to the 
white dwarf radius $R_{wd}$ is only $0.01\:R_{wd}$, which is much smaller 
than the distance from the inner disk edge to the white dwarf (see below).
Only a thin annular region near the inner disk edge is needed to produce the 
variable UV flux.

If our interpretation of the X-ray/UV correlation is correct, the 100-s time
delay can provide an estimate for the inner disk truncation radius.
The radial in-fall velocity in the boundary layer is given by 
$V_r\sim\alpha\:c_s^2 / V_k$ \citep[see e.g.][]{1994ApJ...428L..13N},
where $\alpha$ is the viscosity parameter, $V_k$ the Keplerian velocity,
and $c_s=(kT/\mu m_p)^{1/2}$ the isothermal sound speed.
With the assumption that the temperature in the boundary layer is roughly equal to
the initial temperature of the cooling gas $kT_{max}\approx7\:$keV
(Section \ref{epicspec}), and using $V_k=3200\:$km$\:$s$^{-1}$, 
$\alpha=0.1$, and a mean molecular weight $\mu=0.6$,
we find for the in-fall velocity $V_r\sim35\:$km$\:$s$^{-1}$.
The 100-s time delay then suggests that the inner disk edge is $\sim$0.4$\:R_{wd}$
above the white dwarf.
This result depends on the rather uncertain viscosity parameter $\alpha$
and should only be considered a rough estimate.
However, the inner truncation radius cannot be much larger than our estimate
since this would be inconsistent with the observed orbital modulation of the
X-ray flux (see Section \ref{xraymod}).


\subsection{X-ray luminosity and accretion rate}
\label{lumin}

In quiescence, the combined optical and UV flux from VW Hyi is 
$1.7\times10^{-10}\:\mathrm{erg\:cm^{-2}\:s^{-1}}$, half of which is 
probably due to the accretion disk 
\citep{1987MNRAS.225...73P,1987MNRAS.225..113V}.
For a distance of 65 pc and an inclination of $60^\circ$, the corresponding disk 
luminosity is $L_{disk}=4\times10^{31}\:\mathrm{erg\:s^{-1}}$.
In Section \ref{epicspec} we estimated for the boundary layer luminosity
$L_{bl}=8\times10^{30}\:\mathrm{erg\:s^{-1}}$.
In agreement with earlier observations \citep[e.g.][]{1991A&A...246L..44B}, the 
ratio of boundary layer to disk luminosity $L_{bl}/L_{disk}=0.2$ is well below 
the expected value of 1.
The cause for this lack of boundary layer emission is still an open question
in our understanding of dwarf novae.
For VW Hyi we can rule out a rapidly rotating white dwarf as the cause since 
the rotation velocity of the boundary layer is much smaller than the Keplerian 
velocity (see Section \ref{oscillation}).

As discussed in Section \ref{epicspec}, the X-ray spectrum is in good agreement 
with the model of a multi-temperature plasma having a cooling flow or power-law
emission measure distribution.
This strongly suggests that the origin of the X-ray emission is cooling plasma
in the boundary layer settling onto the white dwarf.
The plasma has an initial temperature of 6--8$\:$keV when it begins to radiate
efficiently in X-rays.
Our results are in remarkable qualitative agreement with the model calculations by
\citet{1993Natur.362..820N} for dwarf novae at low accretion rates.
The calculations predict the presence of a narrow transition region between disk
and boundary layer in which the accreting gas is decelerated well below the
Keplerian velocity.
At this inner disk edge, the gas begins to fall quickly toward the white dwarf and
the density decreases by several orders of magnitude.
Because of the low density, the gas can no longer cool efficiently and becomes
invisible to us.
The accreting material continues to dissipate its rotational kinetic energy while
being heated to temperatures in excess of $10^8$K.
Finally, when the in-falling gas is compressed as it piles up on the white dwarf, 
it begins to cool strongly via bremsstrahlung and becomes visible in X-rays.
This compressed, X-ray emitting gas is localized in a thin region on the white dwarf
and should be well approximated by a simple cooling flow model.
In agreement with our measurements of the boundary layer rotation velocity
(Section \ref{oscillation}), the model calculations predict that the gas in
this thin region has been slowed down to almost the white dwarf rotation velocity.
However, the model does not explain the low ratio $L_{bl}/L_{disk}$.

As the in-falling gas is compressed and cools from an initial temperature
$T_{max}$ to the much lower white dwarf temperature, it releases an enthalpy
$H=5/2\times kT_{max}$ per particle.
We can therefore estimate the rate of accretion from the boundary layer onto the
white dwarf $\dot{M}_{bl}$ using the relation 
\begin{equation}
\label{lbl}
L_{bl}=\frac{5}{2}\:\frac{\dot{M}_{bl}}{\mu m_p}\:kT_{max}
\end{equation}
\citep[e.g.][]{1994ARA&A..32..277F}.
Here $\mu$ is the mean molecular weight (typically $\sim$0.6) and $m_p$ the 
proton mass.
With an initial temperature $kT_{max}\approx7\:$keV and a boundary layer
luminosity $L_{bl}=8\times10^{30}\:\mathrm{erg\:s^{-1}}$,
we find an accretion rate $\dot{M}_{bl}=5\times10^{-12}$M$_\odot\:$yr$^{-1}$.

For comparison, we estimate the accretion rate in the disk $\dot{M}_{disk}$ 
from the disk luminosity $L_{disk}$.
With the reasonable assumption that half of the accretion energy is dissipated 
in the disk, we can calculate $\dot{M}_{disk}$ using the relation
\begin{equation}
\label{ldisk}
L_{disk}=\frac{GM_{wd}\dot{M}_{disk}}{2R_{wd}}
\end{equation}
For a white dwarf mass $M_{wd}=0.63\:$M$_\odot$ and a radius $R_{wd}=8.3\times10^8\:$cm,
the disk accretion rate is $\dot{M}_{disk}=12\times10^{-12}$M$_\odot\:$yr$^{-1}$.
Our result seems to indicate that $\dot{M}_{bl}\approx1/2\:\dot{M}_{disk}$,
i.e. that half of the accreting gas is lost in a wind.
A model for such an outflow in dwarf novae during quiescence is described in
\citet{1994A&A...288..175M}.
However, our estimate of $L_{bl}$ and therefore $\dot{M}_{bl}$ may be too low
if part of the boundary layer is obscured by the disk.
The orbital modulation of the X-ray flux (Section \ref{xraymod}) indicates
that at least some obscuration is present.
A geometrically thin disk can obscure at most half of the boundary layer
so that the actual $\dot{M}_{bl}$ could be up to twice our estimate.
It is therefore still possible that $\dot{M}_{bl}=\dot{M}_{disk}$ and that no
outflow is present.
Our estimate of $\dot{M}_{disk}$ may be too low as well if the disk is truncated
at a radius significantly larger than $R_{wd}$.
In Section \ref{correl} we found that the inner disk edge is probably $\sim$0.4$\:R_{wd}$
above the white dwarf.
This would imply that the actual $\dot{M}_{disk}$ is $\sim$40 per cent larger
than our estimate.
However, it seems unlikely that $\dot{M}_{disk}$ can be sufficiently large so that the low ratio
$L_{bl}/L_{disk}$ can be attributed to a low $\dot{M}_{bl}/\dot{M}_{disk}$.
In order for $\dot{M}_{disk}$ to satisfy $\dot{M}_{bl}/\dot{M}_{disk}=L_{bl}/L_{disk}=0.2$,
the inner disk would have to be truncated at least $\sim$1.0$\:R_{wd}$ above the white dwarf
(or more if we underestimated $\dot{M}_{bl}$).
However, such a large truncation radius is inconsistent with the observed orbital modulation
of the X-ray flux (Section \ref{xraymod}).
We conclude that the low boundary layer luminosity $L_{bl}$ cannot be solely attributed to
a low accretion rate $\dot{M}_{bl}$.

The key to understanding the missing boundary layer emission in VW Hyi may be 
the low initial temperature of the X-ray emitting gas.
If the kinetic energy of the material at the inner disk edge, which is rotating 
at a Keplerian velocity of $\sim$$3200\:$km$\:$s$^{-1}$, is converted completely 
into thermal energy, the plasma would reach a temperature of $\sim$21$\:$keV.
Yet when the in-falling gas first emits X-rays as it piles up on the white dwarf, 
its temperature is only 6--8$\:$keV.
The weak Fe {\sc xxvi} line at 6.9$\:$keV (see Fig. \ref{epic}) clearly demonstrates 
that plasma at higher temperatures is either not present or not radiating 
efficiently in X-rays.
It appears that the gas in the boundary layer looses $\sim$$2/3$ of the available
accretion energy before it begins to emit X-rays.

It is possible that the gas inside the inner disk edge, where it is decelerated 
well below the Keplerian velocity, is cooling so rapidly that it cannot reach 
the maximum temperature of $\sim$21$\:$keV.
The gas would likely have temperatures higher than the inner disk but not high 
enough to emit X-rays that could be detected by \xmm.
However, no strong EUV emission from VW Hyi is observed, and, owing to the low 
hydrogen column density, not much of the EUV flux can be hidden by interstellar 
absorption.
From the non-detection of EUV emission at wavelength shorter than 98 nm 
\citep{1996ApJ...466..964L} and from the quiescent flux detected by EXOSAT 
\citep{1996A&A...307..137W}, we estimate that at most 20 per cent of the 
expected boundary layer luminosity is emitted in the EUV.

\citet{1992MNRAS.259P..23L} found that a weak magnetic field ($\sim$$10^4\:$G)
of the white dwarf can truncate the inner disk, thus causing the UV delay
observed during outburst in some dwarf novae including VW Hyi.
The density inside the inner disk edge may be sufficiently low for the magnetic 
field to pentrate into the boundary layer and cause the plasma to emit cyclotron 
radiation.
We estimate the cyclotron cooling time scale $t_{cycl}$ for thermalized,
non-relativistic electrons from the emitted cyclotron power
\citep[equ. 6.7 in][]{1979rpa..book.....R} and find
$t_{cycl}\left[\mathrm{s}\right]=4\times10^8\times B\left[\mathrm{G}\right]^{-2}$.
For a magnetic field $B=10^4\:$G, the cyclotron cooling time is $\sim$4$\:$s,
comparable to the in-fall time of gas not supported by rotation.
A significant fraction of the thermal energy could be lost to cyclotron 
radiation that would be emitted in the radio band.
However, strong radio emission from dwarf novae is not observed 
\citep[e.g.][]{1989A&A...218..137B}.
Also, we did not detect strictly periodic X-ray oscillations that would indicate 
emission from a magnetic pole.

\citet{1994A&A...288..175M} suggested that during quiescence the inner accretion 
disk evaporates and forms a corona.
The gas in the corona is partially accreted onto the white dwarf and partially
lost in a wind.
The X-ray spectrum is dominated by emission from the cooling plasma piling up on 
the white dwarf.
In this model, some of the thermal energy in the corona is lost to the wind or 
conducted back to the accretion disk, which may explain the low temperature of 
the plasma when it begins to emit X-rays.


\subsection{Orbital X-ray modulation}
\label{xraymod}

Our investigation of the orbital X-ray modulation showed that the spectrum
becomes softer when the flux is decreasing (Section \ref{epicspec}).
This spectral change is opposite of what would be expected if the decline in
X-ray flux were due to absorption by some intervening gas
(e.g. an extended bright spot).
We therefore suggest that the orbital X-ray modulation is caused by a geometric 
effect, i.e. that during the faint phase some hot regions of the boundary 
layer are blocked from view, while cooler regions remain visible.
For this to be possible, the obscuring material has to be comparable in size and 
close to the boundary layer (e.g. the white dwarf or the inner accretion disk).
We consider three possible explanations:

\begin{enumerate}
\item
The boundary layer may not be axisymmetric.
It is difficult to imagine how an asymmetry that is synchronized with the orbital 
motion can exist on the rapidly rotating white dwarf.
One possibility is asymmetric accretion caused by a non-axisymmetric inner disk.
Again, it is unclear how such an asymmetry in the disk could persist for several 
orbital cycles.
In some dwarf novae, stationary structures have been observed in the form of 
spiral waves \citep[e.g.][]{2001astr.conf...45S}, albeit only during outbursts 
and in the outer parts of the disk.
\item
The boundary layer may be partially obscured by a non-axisymmetric inner disk.
Since the transition from optically thick disk to optically thin boundary 
layer is quite abrupt \citep{1993Natur.362..820N}, the disk essentially acts 
like a circular aperture, blocking part of the X-ray emission.
The fraction of visible boundary layer should remain constant, yet it may vary 
with orbital phase if the inner edge of the disk is not circular.
Again, it is unclear what the cause of such an asymmetry might be.
\item
The rotation axes of the white dwarf and the inner disk may not be aligned.
It is thought that the boundary layer forms a belt-like structure on the white 
dwarf.
Since the X-ray emitting part of the boundary layer is mostly co-rotating with 
the white dwarf, it is likely aligned with the equator.
If the inner disk is tilted with respect to the white dwarf's equator, the 
fraction of boundary layer obscured by the disk may vary with phase, causing an 
orbital modulation of the X-ray flux.
\end{enumerate}

In the latter scenario, it is unlikely that the rotation axis of the white dwarf 
is tilted with respect to the orbital plane since any such misalignment would 
disappear on a time scale of 100 years \citep{1978assl...68.....K}.
However, warping of the accretion disk may cause a tilt of the inner disk edge.
\citet{1998MNRAS.300..561M} found that tidal interactions are too weak to cause 
warping during the high state, but a tilt may be present in the quiescent disk 
of short-period dwarf novae.
\citet{1999ApJ...524.1030L} suggested that warping of the inner disk may be 
caused by a weak magnetic field of the primary.
The precession of a warped disk in VW Hyi could be revealed if future 
observations show a phase shift of the orbital X-ray modulation relative to the 
orbital hump seen in the UV.


\section{Conclusions}

We have analyzed \xmm\ data of the dwarf nova VW Hyi obtained during quiescence.
The spectrum indicates that the X-ray emission is due to an optically thin 
plasma that cools as it settles onto the white dwarf.
The cooling plasma has an initial temperature of 6--8$\:$keV when it first 
emits X-rays.
This temperature is well below the $\sim$21$\:$keV expected if all the rotational 
kinetic energy of the material in the inner disk were instantly converted into 
heat.
Apparently, the accreting gas looses $\sim$$2/3$ of its thermal energy before
it piles up on the white dwarf and begins to radiate efficiently in X-rays.
We estimated for the luminosity of the boundary layer
$8\times10^{30}\:\mathrm{erg\:s^{-1}}$, which is only 20 per cent of the 
accretion disk luminosity.
The low boundary layer luminosity cannot be explained by a low mass transfer 
rate.
We derived for the accretion rate onto the white dwarf
$5\times10^{-12}$M$_\odot\:$yr$^{-1}$ and for that in the disk
$12\times10^{-12}$M$_\odot\:$yr$^{-1}$.
We determined elemental abundances for some common elements and found no large 
deviations from the solar values.

From the Doppler-broadening of emission lines, we determined that the X-ray 
emitting part of the boundary layer is rotating with a velocity 
$V_{bl}\:\sin\:i=540\:$km$\:$s$^{-1}$, which is similar to the white dwarf 
rotation velocity but much smaller than the Keplerian velocity of the inner 
disk.
This result rules out a rapidly rotating white dwarf as the cause of the low 
boundary layer luminosity.
We found a 60-s quasi-periodic X-ray oscillation with a coherence time of only
$\sim$100 s.
The oscillation is likely due to the rotation of the boundary layer.
The corresponding rotation velocity $V_{bl}\:\sin\:i=750\:$km$\:$s$^{-1}$ 
is consistent with that derived from the line-broadening.

The low-frequency variability of the X-ray and the UV flux is well described
by a power-law spectral density function with an index consistent with a
random walk noise process ($\beta=2$).
In X-rays the variability is caused by fluctuations of the accretion rate
onto the white dwarf.
We found a correlation between the X-ray and the UV variability with the X-ray
fluctuations being delayed by $\sim$100 s.
This result suggests that the variable UV flux is originating from near the transition 
region between disk and boundary layer and that accretion rate fluctuations in 
this region are propagated to the X-ray emitting part of the boundary layer 
within $\sim$100 s.
We estimated from the 100-s time delay that the accretion disk is truncated
$\sim$0.4$\:R_{wd}$ above the white dwarf.
An orbital modulation of the X-ray flux suggests that the inner accretion disk 
is tilted relative to the white dwarf's equator and the orbital plane.
This warping of the inner disk may be due to a weak magnetic field of the white 
dwarf.


\section*{acknowledgements}

This work is based on observations obtained with \xmm, an ESA science mission 
with instruments and contributions directly funded by ESA Member States and the 
USA (NASA).
The authors acknowledge support from NASA grants NAG5-7714 and NAG5-12390.


\bibliographystyle{mn2e}
\bibliography{vwhyi}

\label{lastpage}

\end{document}